# Semiconductor quantum dots with light-hole exciton ground state: fabrication and fine structure

Y. H. Huo,[1*] B. J. Witek,[2] S. Kumar,[1] R. Singh,[3] E. Zallo,[1] R. Grifone,[4] D. Kriegner,[4] R. Trotta,[1,4] N. Akopian,[2] J. Stangl,[4] V. Zwiller,[2] G. Bester,[3*] A. Rastelli,[1,4*] and O. G. Schmidt,[1]

[1]*Institute for Integrative Nanosciences, IFW Dresden, Helmholtzstraße 20, D-01069 Dresden, Germany*

[2]*Kavli Institute of Nanosciene, Delft University of Technology, Lorentzweg 1, 2628CJ Delft, The Netherlands*

[3]*Max-Planck-Institute for Solid State Research, Heisenbergstraße 1*

*70569 Stuttgart, Germany*

[4]*Institute of Semiconductor and Solid State Physics, Johannes Kepler University Linz, Altenbergerstraße 69, A-4040 Linz, Austria*

Quantum dots (QDs) can act as convenient hosts of two-level quantum systems, such as single electron spins, hole spins or excitons (bound electron-hole pairs). Due to quantum confinement, the ground state of a single hole confined in a QD usually has dominant heavy-hole (HH) character. For this reason light-hole (LH) states have been largely neglected, despite the fact that may enable the realization of coherent photon-to-spin converters or allow for faster spin manipulation compared to HH states. In this work, we use tensile strains larger than 0.3% to switch the ground state of excitons confined in high quality GaAs/AlGaAs QDs from the conventional HH- to LH-type. The LH-exciton fine structure is characterized by two in-plane-polarized lines and, ~400 μeV above them, by an additional line with pronounced out-of-plane oscillator strength, consistent with theoretical predictions based on atomistic empirical pseudopotential calculations and a simple mesoscopic model.

*electronic mails: y.huo@ifw-dresden.de; armando.rastelli@jku.at; g.bester@fkf.mpg.de.

Epitaxial semiconductor quantum dots (QDs) are considered as candidate building blocks for quantum technologies, as they can act both as hosts of static quantum bits (excitons [1, 2] or spins [3-13]) or as triggered sources of single and entangled photons [14-16]. In particular, QDs can confine carriers whose spin is characterized by longer coherence times compared to spins in bulk. In III-V semiconductors, hole spins are receiving increasing attention, as decoherence due to hyperfine interaction should be reduced compared to electron spins [8-13]. All experimental studies presented so far have been dealing with heavy-holes (HH). This is because quantum confinement lifts the valence band degeneracy and leads to states with HH character which are energetically well above the light-hole (LH) states. Further energetic separation is provided by the compressive strain, which usually characterizes self-assembled QDs such as InGaAs QDs inside a GaAs matrix. Some proposals suggest, however, that using LHs instead of HHs would be beneficial for quantum information technologies. These include the coherent conversion of photons into electron-spins [17], faster qubit control operations compared with HH [18, 19], the possibility to control the LH spin state via microwave or RF signals, and the efficient control of a magnetic impurity spin coupled to a QD [19, 20]. It is important to note that for these proposals to be used, the LH should be the ground state (GS), otherwise decay to GS would limit its coherence time [21]. While QD excitons involving an excited-state hole with LH character have been studied [22, 23], reports on QDs with LH GS are limited to Ref. [24]. However, the broad emission linewidth of these QDs prevented access to the excitonic fine structure and a detailed study of LH excitons. From a theoretical point of view, the fine structure of QDs with LH-exciton GS has not been calculated so far.

Here we present a method to realize high-quality epitaxial QDs with GS of dominant LH-type and we discuss experimentally and theoretically the fine structure of their excitonic emission. The QDs are initially unstrained GaAs QDs in an AlGaAs matrix, which possess a GS with dominant HH character, but are tall enough to allow the LH to become the GS with moderate tensile strains. In-plane biaxial tensile strains larger than 0.3% are induced onto the QDs by embedding them into symmetrically prestressed membranes, which are then released from the substrate, see Fig. 1(b). (The strain is defined here as $\varepsilon=(a_s-a_0)/a_0$ where $a_s$ is the lattice constant of the stressed structure and $a_0$ is the lattice constant of the epitaxial layer prior to release.) Additional fine-tuning of the biaxial strain is achieved by placing the membranes onto a piezoelectric actuator. Polarization-resolved micro-photoluminescence (μ-PL) spectroscopy measurements show that the GS excitonic emission is characterized by three lines, one of which is mostly polarized along the growth direction and is separated from the others by a very large splitting of several hundred µeV, in good agreement with our theoretical results. µ-PL measurements in magnetic field allow also the dark excitons to be observed and further demonstrate the achievement of three-dimensionally confined excitons with LH GS.

The QDs studied here were obtained by local droplet etching of nanoholes into a GaAs surface [25-27] followed by heterostructure overgrowth [28-29]. The advantage of this method over strain-induced self-assembly is that the height of the QDs, which directly influences the LH-HH splitting, can be controlled in a wide range even when the depth of the holes is fixed [30]. In addition, these QDs can display single-dot emission linewidths narrower than 25 µeV [28], allowing detailed investigations of the excitonic fine-structure to be performed. For more details, see methods. To gather information on the dot morphology, we measured by atomic force microscopy (AFM) the surface morphology of two additional samples where the growth was interrupted after the deposition of the bottom barrier and after GaAs overgrowth, respectively. AFM images of two similarly shaped nanoholes found on the two samples are shown on the left side of Fig. 1(a). The approximate shape of the GaAs QDs can be obtained by subtracting the two images, as also shown in Fig. 1(a). Linescans of the QD interfaces as well as of the resulting QD shape along two orthogonal directions are shown on the right part of Fig. 1(a). We see the GaAs-filled nanohole has an irregular shape, which is elongated in the [110] direction. Similar conclusions can be drawn by inspection of other nanoholes.

To generate tensile strain into the GaAs QDs, we follow an approach similar to Refs. [32-34]. The QDs were placed into symmetrically pre-stressed membranes which include one $In_{0.2}Al_{0.3}Ga_{0.5}As$ stressor layer below and another above the active structure (see Fig. 1(b) left). After defining patterns by optical lithography and wet chemical etching, we selectively remove a sacrificial $Al_{0.75}Ga_{0.25}As$ layer placed below the membrane structure. In this way the membranes are undercut and bond-back to the underlying substrate, as shown in the sketch on the right of Fig. 1 (b). After release, the strain, which was originally confined into the InAlGaAs layers, is shared also with the initially unstrained heterostructure (see horizontal arrows in Fig. 1(b), with lengths proportional to the strain magnitudes).

Strain was quantified by X-ray diffraction measurements of a similar sample containing a GaAs layer instead of the (Al,Ga)As heterostructure between the $In_{0.2}Al_{0.3}Ga_{0.5}As$ layers. X-ray diffraction scans along the growth direction around the (004) Bragg peak are shown in Fig. 1(c) in red together with simulation results based on dynamical scattering theory (blue). See Methods for details. From the position of the peak associated to strained GaAs we obtain a strain value of 0.36%, which is close to the expected geometrical average of 0.38% [34] and is consistent with the red-shift of the free-exciton emission of the GaAs layer after undercut. The effect of tensile strain on the QD emission can be seen in the ensemble PL spectra of Fig. 1(d), which were collected from a membrane before (blue line) and after (red line) the removal of the sacrificial layer. After undercut, the average QD emission is red-shifted by about 40 meV.

Figure 2 illustrates the effect of strain on the emission of single QDs, which was excited and collected both along the [001] growth direction and perpendicular to it, i.e. along the [1-10] crystal direction. The spectra of as-grown QDs collected at low excitation-power are characterized by a neutral exciton emission which is composed of two linearly polarized lines $A_x$ and $A_y$ separated on average by a fine-structure-splitting (FSS) of 60 µeV (see Fig. 2(a)). The high energy component $A_x$ is aligned on average along the [1-10] crystal direction, consistent with the elongation of the QDs (see Fig. 1(a) and Ref. [36]). As expected, the spectra collected from the sample side show mainly a single line, which is linearly polarized roughly along the [110] crystal direction, as shown for a different QD in Fig. 2(b). Similar results were reported previously for conventional InGaAs QDs [37].

The spectra dramatically change after undercutting, as seen in Figs. 2(c) and (d). The average FSS of the doublet A decreases from 60 to 13 µeV, $A_x$ and $A_y$ also swapped their energy order (compare Figs. 2(a) and (c)) and the low energy component $A_x$ (polarized on average along [1-10]) becomes significantly weaker than $A_y$. Most importantly, an additional new line, labeled as $B_x$ in Fig. 2(c), appears on the high energy side of A. This weak line is linearly polarized on average along the [1-10] direction and is energetically separated by $\Delta E \sim 420$ µeV from $A_x$. The ratio between the intensity of $B_x$ and $A_y$ was found to be independent on excitation power indicating an equal generation rate for the excitonic states A and B. When observed from the cleaved edges, we see that the B line ($B_z$ in Fig. 2(d)) becomes very prominent and is linearly polarized along the growth direction. The above observations hold for several tens of QDs investigated so far.

To understand these findings qualitatively, we first derived from the theory of invariants [38] the spin-spin interaction Hamiltonian and solve it for a pure heavy- and a pure light-hole exciton. For a pure HH exciton, by assuming that the momentum matrix elements are isotropic, $\Pi = \langle s|p_x|X\rangle = \langle s|p_y|Y\rangle = \langle s|p_z|Z\rangle$, we find for the transition dipoles of the four excitonic states:

$$|4\rangle = 0, |3\rangle = \Pi e_x, |2\rangle = -i\Pi e_y, |1\rangle = 0,$$

where $e_x$, and $e_y$ are unit vectors along the [110] and [1-10] directions. There are two bright states which are in-plane polarized along the [110] and [1-10] directions and two dark states. For a pure LH exciton the transition dipoles of the four excitons become:

$$|4\rangle = \frac{\Pi e_x}{\sqrt{3}}, |3\rangle = \frac{-i\Pi e_y}{\sqrt{3}}, |2\rangle = \frac{-2\Pi e_z}{\sqrt{3}}, |1\rangle = 0$$

In contrast to the HH exciton case, three exciton states are bright: two polarized in-plane along [110] and [1-10], and one polarized along the [001] growth direction. This model transparently shows the qualitative expectations in terms of polarizations, but fails to give the energetic splitting between the states that is required to convincingly interpret the experimental results.

To obtain more quantitative results, we have performed numerical calculations of the exciton states, using the atomistic empirical pseudopotential method and configuration interaction for up to three million atoms [39]. We calculate the excitonic states, including the electron-hole exchange interaction [40] for tall lens-shaped GaAs QDs with a height of 8 nm and diameters varying between 15 and 35 nm, embedded in $Al_{0.3}Ga_{0.7}As$. The structure is nearly strain-free and due to the rather large height, the splitting between the dominant HH and dominant LH states is significantly reduced, compared to the common InGaAs/GaAs case. Figure 3(a) shows the HH and LH character of the first six hole states in an unstressed GaAs QD with 35 nm diameter and 8 nm height. The first hole state $h_0$ has over 90% HH character, but the deeper holes, which are only a few meV from $h_0$ have already a significant LH character. The dependence of the hole admixture is non-monotonic, since the amount of LH mixing depends on the orbital symmetry of the states. To create a ground state hole ($h_0$) with dominant LH character, we apply tensile strain to the simulation cell. Under the stressed conditions, all the atomic positions are relaxed to minimize the strain energy, using the valence force field method [39]. In Fig. 3(b) we plot the HH and LH character of the first hole state $h_0$ as a function of the induced strain. We see that, already for rather moderate tensile strain values of about 0.1%, the character shifts from dominantly HH to dominantly LH. In Fig. 3(c) to (e) we plot the excitonic fine structure as a function of the tensile strain for three different QDs. The size of the symbols is proportional to the oscillator strength of the transition. The energies are given relative to the lowest exciton states that is, in all cases, a dark state. The color of the symbols gives the polarization property of the transitions, red for in-plane polarization and green for out-of-plane (along [001]) polarization. At zero strain we have the well-known situation where the two bright states are polarized along the [110] and [1-10] directions and are split by the FSS (17.5, 14.8, 10.7 µeV for the QD with d=15, 20, 25 nm, respectively), while the dark states are nearly degenerate. The bright-dark splittings progressively reduces as well (140, 104, 80 µeV) as the diameter increases (15, 20, 25 nm). When tensile strain is introduced, the situation changes dramatically with one bright state with [001] polarization between 200 and 500 µeV above the lowest dark state and two bright states, split by only a few µeV (0-10 µeV) and polarized in-plane, at an energy of around 30 µeV (46, 37, 30 µeV for d= 15, 20, 25 nm, respectively at a strain value of 0.6%) above the lowest dark state. These results are in agreement with the mesoscopic model that predicted transitions with these polarizations.

Although the modeled structure is idealized, we can compare the calculation results to the experiment. We first note that the FSS of the in-plane polarized lines, which are labeled as $A_x$ and $A_y$ drops both in theory and experiment as strain is introduced (see Fig. 2(a-d) and red points in 3(c-e)). The physical origin of this drop is not yet clear. Most importantly, a new z-polarized line (red symbols in Fig. 3(c-e)) appears, which we associate to the line B seen in the experiment. Its energy separation ΔE from the in-plane polarized doublet for a strain of about 0.35% varies between 250 and 450 µeV depending on the model size and is thus comparable with the experimental values (see Fig. 2(c-d)). While in the experiment we cannot draw conclusions about possible dependencies of ΔE on dot size, we can test the predicted increase of ΔE for increasing tensile strain (see Fig. 2(e)). To this aim, prestrained membranes were transferred onto a piezoelectric substrate via gold-thermocompression bonding (see [41-43] and inset of Fig. 2(e)), allowing us to increase (decrease) the tensile strain by simply decreasing (increasing) the electric field applied across the piezo. When the electric field changes from 23.3 to -10 kV/cm, the exciton energies decrease by about 9 meV and ΔE increases from about 474 to 518 µeV in agreement with the predicted trend.

Finally, measurements in magnetic field allow us to access the dark excitonic states and provide further evidence of the switching of the valence-band GS from the HH- to the LH-type. PL spectra of a single QD in the as-grown sample are shown in Fig. 4(a) and reveal a typical behavior of a HH exciton in the Faraday configuration [43]. We notice that for fields larger than about 4 T a weak emission from the initially dark excitons appears, which we ascribe to the pronounced shape anisotropy of the QDs (see the following). After subtraction of the diamagnetic shift [31], the exciton state energies versus magnetic field are plotted in Fig. 4(b). The experimental data points can be fitted using a Hamiltonian consisting of the exchange interaction and magnetic-field dependent Zeeman part. As fitting parameters, we obtain the electron and hole g-factors ($g_{e,z}$ = 0.90, $g_{h,z}$ = 0.04), and the exchange energies ($\delta_0$ = 140 µeV, $\delta_1$ = 38 µeV, $\delta_2$ = 9 µeV).

In case of the undercut membrane in Fig. 4(c) we observe a remarkably different picture. All four exciton transitions are clearly visible due to state mixing in magnetic field. In particular an additional line, which we attribute to the initially dark state ($B_{dark}$) can be identified. If we extrapolate the position of the dark ground state $B_{dark}$ to zero magnetic field, we obtain a $B_{dark}$ – $B_x$ splitting (398 µeV) as large as predicted for a LH exciton. This enormous exchange energy dominates the Zeeman term. As a result, the dependence of the energy splitting on the magnetic field deviates from a linear behavior. In addition, the energy separation between $B_{dark}$ and the A-doublet is substantially reduced as compared to the energy separation of the dark excitons and the A-doublet in the unstrained QDs, consistent with the theory prediction (see Fig. 3).

In spite of the overall good agreement between theory and experiment, further work is required to understand the relative intensities and polarization properties of the observed lines. As an example the B line should not be observable when measured along the [001] direction, according to the model. We qualitatively ascribe this observation to the pronounced anisotropy of the dot shape (see Fig. 1(a)). In particular we expect the broken inversion symmetry along the [1-10] direction to affect the relative intensities and polarization properties of the emitted lines by inducing LH-HH mixing. The pronounced polarization anisotropy of the A doublet in the membrane sample (see Fig. 2(c)) is a clear manifestations of residual hole mixing. (Note that the polarization anisotropy of the in-plane polarized lines for an exciton of dominant LH-type is much more sensitive to mixing than a dominantly-HH exciton, see eq. (7) of Ref. [44]). The implementation of more realistic structures into the numerical calculations shall allow us to unveil the impact of asymmetries in the light emission of excitons with dominant LH character.

In conclusion, we have shown that the excitonic ground-state of self-assembled GaAs QDs can be switched from the common dominant heavy-hole type to light-hole type by releasing prestressed membranes with initially unstrained QDs. The excitonic fine structure was investigated both experimentally and theoretically. The high optical quality of the presented dots, the compatibility of membrane processing with electrical control [42,43], and the prediction that LH-HH mixing as low as 5% can be achieved for realistic strains of 0.4%, demonstrate that three-dimensionally confined light-holes can soon be explored as new semiconductor-based quantum systems for quantum communication technologies.

We acknowledge P. Atkinson, Ch. Deneke, D. Thurmer for assistance with the MBE and G. Katsaros, and R. Rezaev for fruitful discussions. This work was financially supported by the BMBF project QuaHL-Rep (Contracts no. 01BQ1032 and 01BQ1034) and the DFG FOR730.

**Methods**

**Sample growth**

The samples studied here were grown by molecular beam epitaxy (MBE) on semi-insulating GaAs(001) substrates. For this work, 16-nm deep nanoholes were obtained by depositing 11.4 monolayers (ML) of excess Ga on a GaAs(001) surface at a substrate temperature of $515^{o}C$ followed by 5 min annealing under As flux. The nanoholes were then overgrown with 7 nm $Al_{0.44}Ga_{0.56}As$ as bottom barrier, followed by 2.15 nm GaAs (QD material), 2 min annealing favoring hole filling, 25 nm $Al_{0.37}Ga_{0.63}As$, and 10 nm graded $Al_xGa_{1-x}As$ (with x varying from 0.37 to 0.44 by changing the supplied flux of Al).

**Strain characterization via X-ray diffraction**

The XRD data were collected using a Panalytical X-Pert Pro diffractometer with sealed Cu-tube and a hybrid monochromator consisting of a parabolic multilayer mirror and a Ge(220) channel cut monochromator. The optics was adjusted to CuKα1 radiation and the beam was cut to (1x2)mm by slits. The (004) Bragg reflection is sensitive to the lattice parameters in growth direction, which change in the opposite direction compared to the in-plane lattice parameter upon layer undercut. Before the undercut, signals from the $In_{0.2}Al_{0.3}Ga_{0.5}As$ and the sacrificial AlGaAs layers are observed in Fig. 1(c), while the GaAs layer placed between the InAlGaAs layers has the same lattice parameter as the GaAs substrate, so that no extra peak shows up. Oscillations in the scans are due to the finite layer thicknesses along the [001] direction. After undercut, the peak due to InAlGaAs shifts to higher Q values, corresponding to smaller lattice parameters in growth direction. This is consistent with an increase of the in-plane lattice parameter of this layer. The GaAs layer is strained accordingly, and its lattice parameter in growth direction decreases as well, giving rise to the additional peak at the right side of the GaAs substrate peak.

**Optical characterization**

The optical properties of the structures were characterized by µ-PL spectroscopy at temperatures below 10 K. PL was excited with a laser wavelength of 532 nm using a microscope objective with numerical aperture of 0.42 and spectrally analyzed with a resolution of about 20 µeV. The linear polarization content of the emitted light was analyzed by a rotatable achromatic half-wave plate retarder followed by a fixed polarizer in the collection path. For side collection, samples were carefully cleaved along the [110] direction and glued vertically to a copper holder (see inset in Fig. 2(b)). For polarization-resolved measurements, samples were aligned in such a way that a polarization angle of 0 degrees corresponds to the [110] crystal direction within about 5°. Magneto-PL measurements were performed in the Faraday configuration for fields up to 8 T using an objective of 0.85 numerical aperture.

**Figures:**

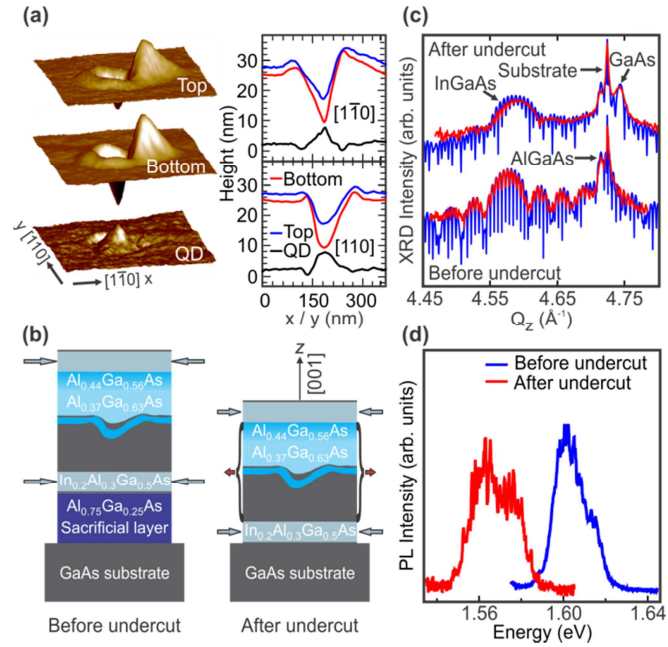

FIG. 1

(color online). (a) Representative AFM images of the surface of GaAs-filled AlGaAs nanoholes (Top) and of the AlGaAs nanoholes (Bottom). The difference between the two images provides the approximate morphology of the resulting GaAs QD. Linescans of the AFM images along [1-10] and [110] crystal directions are shown on the right side. (b) Side-view sketches of as grown sample structure before (left) and after removal of the sacrificial layer (right). The length of the horizontal arrows is proportional to the magnitude of in-plane strain in the layers. (c) X-ray diffraction radial (00L) scan around the (004) Bragg peak of the sample before (bottom) and after (top) removal of the sacrificial layer, measured data in red, simulated curves in blue, respectievly. (d) Ensemble PL spectra of QDs before (top) and after (bottom) undercut.

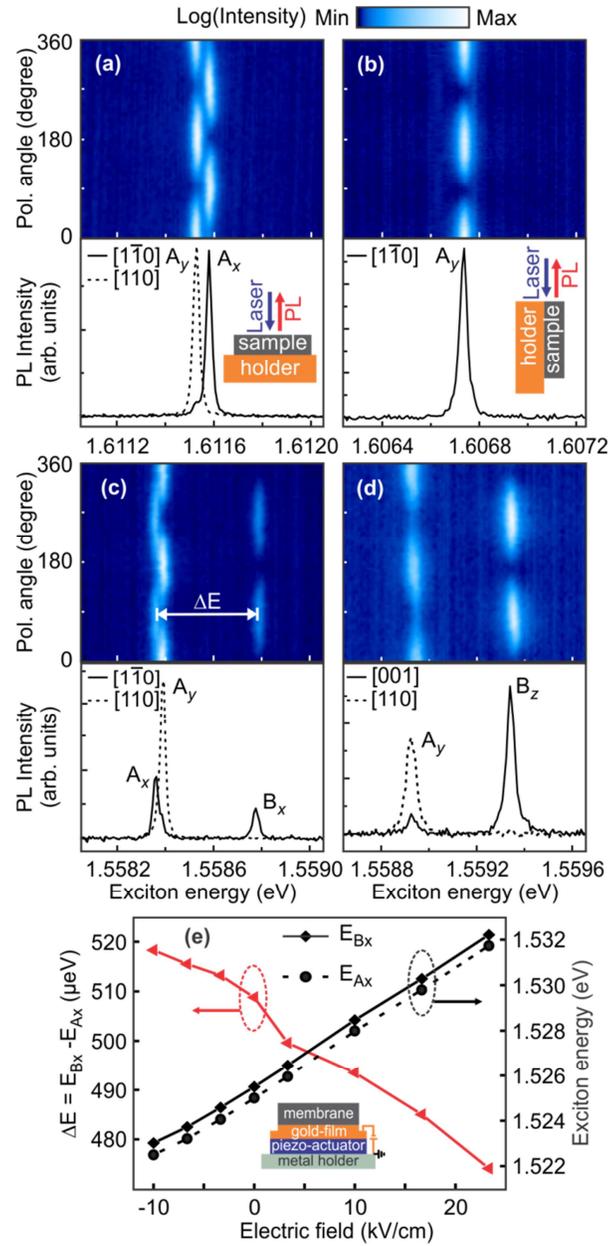

FIG. 2 (color on line). Polarization-resolved µPL spectra of representative QDs collected along the growth direction (a and c, see sketch in the inset of a) or along the [1-10] crystal direction (b and d, see sketch in the inset of b) from the as-grown sample (a and b) or from released and bonded-back membranes (c and d). In (a and c) linear-polarization angles of 0 and 90° correspond to the [110] and [1-10] directions, respectively. In (b and d) angles of 0 and 90° correspond to the [110] and [001] directions, respectively. (e) Dynamic strain tuning of exciton energy and energy splitting ΔE for a QD in a membrane which was released from the GaAs substrate and placed onto a piezoelectric actuator. The inset is a sketch of the experimental configuration.

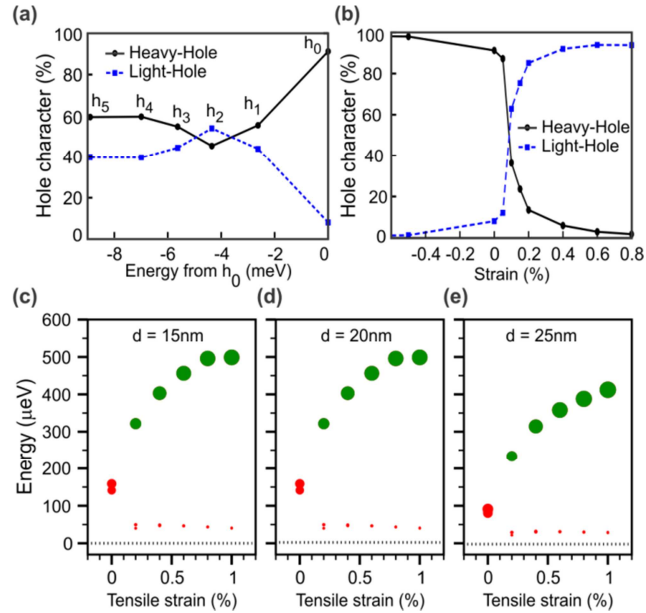

FIG. 3 (color online). (a) Analysis of the hole character of the first six hole states in an unstrained lens-shaped GaAs QD with 35 nm base diameter and 8 nm height. The energy is given with respect to the energy of the first hole state $h_0$. (b) Analysis of the hole character of the first hole state $h_0$ in a lens-shaped GaAs QDs with 35 nm base diameter and 8 nm height as a function of strain. (c) to (e) Excitonic fine structure for GaAs QDs with varying diameter d and a constant height of 8 nm, as a function of the tensile strain. The energies are given by setting the energy of lowest exciton state (dotted line) to zero. The red circles represent the transitions polarized in the growth plane (001) while the green circles represents the transitions polarized along the growth direction [001]. The size of the circles is proportional to the oscillator strengths. The lowest exciton transition energies are 1.645 eV, 1.629 eV, 1.619 eV, 1.608 eV for d = 15, 20, 25, 35 nm, respectively.

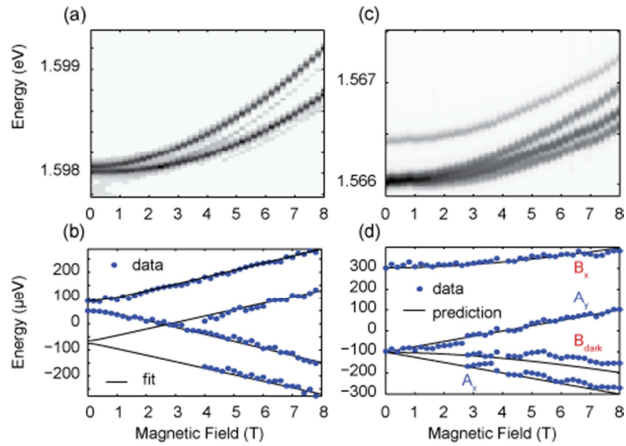

FIG. 4 (color online). µ-PL spectra of two QDs taken as a function of magnetic field applied along the [001] direction (Faraday configuration) on the as grown sample (a) and in the undercut membrane (c). (b) and (d) show the experimental data points of the Zeeman splitting after subtraction of the diamagnetic shifts.

*References*

1. Zrenner, A. *et al.* Coherent properties of a two-level system based on a quantum-dot photodiode. Nature **418,** 612-614 (2002).

2. Ramsay, A. J. A review of the coherent optical control of the exciton and spin states of semiconductor quantum dots. Semicon. Sci. Technol. **25,** 103001 (2010).

3. Greilich, A. *et al.* Mode Locking of Electron Spin Coherences in Singly Charged Quantum Dots. Science **313,** 341-345 (2006).

4. Press, D., Ladd, T.D., Zhang, B., & Yamamoto, Y. Complete quantum control of a single quantum dot spin using ultrafast optical pulses. Nature **456,** 218-221 (2008).

5. Khaetskii, A. V. & Nazarov, Y. V. Spin relaxation in semiconductor quantum dots. Phys. Rev B **61,** 12639-12642 (2000).

6. Loss, D. & Divincenzo, D. P. Quantum computation with quantum dots. Phys. Rev. A, **57,** 120-126 (1998).

7. Nowack, K. C., Koppens, F. H. L., Nazarov, Y. V. & Vandersypen, L. M. K. Coherent control of a single electron spin with electric fields. Science **318,** 1430-1433 (2007).

8. Laurent, S. *et al*. Electrical control of hole spin relaxation in charge tunable InAs/GaAs quantum dots. Phys. Rev. Lett. **94,** 147401 (2005).

9. Gerardot, B. D. *et al*. Optical pumping of a single hole spin in a quantum dot. Nature **451,** 441-444 (2008).

10. Brunner, D. *et al.* A coherent single-hole spin in a semiconductor. Science **325,** 70-72 (2009).

11. Fallahi, P., Yılmaz, S. T. & Imamoğlu, A. Measurement of a heavy-hole hyperfine interaction in InGaAs quantum dots using resonance fluorescence. Phys. Rev. Lett. **105,** 257402 (2010).

12. Greve, K. D. *et al.* Ultrafast coherent control and suppressed nuclear feedback of a single quantum dot hole qubit. Nature Physics, **7,** 872-878 (2011).

13. Yamamoto, Y. *et al*. Optically controlled semiconductor spin qubits for quantum information processing. Phys. Scr. T137, 014010 (2009).

14. Michler, P. *et al*. A quantum dot single-photon turnstile device. Science 290, 2282-2285 (2000).

15. Akopian, N. *et al*. Entangled photon pairs from semiconductor quantum dots. Phys. Rev. Lett. **96**, 130501 (2006).

16. Salter, C. L. *et al*. An entangled-light-emitting diode. Nature **465,** 594-597 (2010).

17. Vrijen, R. & Yablonovitch, E. A spin-coherent semiconductor photo-detector for quantum communication. Physica E **10,** 569-575 (2001).

18. Sleiter, D. & Brinkman, W. F. Using holes in GaAs as qubits : An estimate of the Rabi frequency in the presence of an external rf field. Phys. Rev. B **74,** 153312 (2006).

19. Reiter, D. E., Kuhn, T. & Axt, V. M. All-optical spin manipulation of a single manganese atom in a quantum dot. Phys. Rev. Lett. **102,** 177403 (2009).

20. Reiter, D. E., Kuhn, T. & Axt, V. M. coherent control of a single Mn spin in a quantum dot via optical manipulation of the light hole exciton. Phys. Rev. B **83,** 155322 (2011).

21. Schmidt, K. H., Medeiros-Ribeiro, G., Oestreich, M., Petroff, P. M., & Döhler, G. H. Carrier relaxation and electronic structure in InAs self-assembled quantum dots. Phys. Rev. B, **54,** 11346-11353 (1996).

22. Karlsson, K. F. *et al*. Optical polarization anisotropy and hole states in pyramidal quantum dots. Appl. Phys. Lett. **89,** 251113 (2006).

23. Karlsson, K. F. *et al*. Fine structure of exciton complexes in high-symmetry quantum dots: effects of symmetry breaking and symmetry elevation. Phys. Rev. B **81,** 161307(R) (2010).

24. Troncale, V., karlsson, K. F., Pelucchi, E., Rudra, A. & Kapon. E. Control of valence band states in pyramidal quantum dot-in-dot semiconductor heterostructures. Appl. Phys. Lett. **91,** 241909 (2007).

25. Lee, J. H., Wang, Zh. M., Strom, N. W., Mazur, Yu. I. & Salamo, G. J. InGaAs quantum dot molecules around self-assembled GaAs nanomound templates. Appl. Phys. Lett. 89, 202101 (2006).

26. Heyn, C., Stemmann, A. & Hansen, W. nanohole formation on AlGaAs surfaces by local droplet etching with gallium. Journal of Crystal Growth **311,** 1839-1842 (2009).

27. Zallo, E., Atkinson, P., Rastelli, A. & Schmidt, O. G. Controlling the formation of quantum dot pairs using nanohole templates. Journal of Crystal Growth **338,** 232-238 (2012).

28. Kumar, S. *et al.* Strain-induced tuning of the emission wavelength of high quality GaAs/AlGaAs quantum dots in the spectral range of the $^{87}$Rb $D_2$ lines. Appl. Phys. Lett. **99,** 161118 (2012).

29. Atkinson, P., Zallo, E. & Schmidt, O.G. Independent wavelength and density control of uniform GaAs/AlGaAs quantum dots grown by infilling self-assembled nanoholes. J. Appl. Phys. (in press)

30. Wang, L. *et al.* Self-assembled quantum dots with tunable thickness of the wetting layer: role of vertical confinement on interlevel spacing. Phys. Rev. B **80,** 085309 (2009)

31. Supplementary information on experiment and theory is available on request.

32. Cohen, G. M., Monney, P. M., Paruchuri, V. K. & Hovel, H. J. Dislocation-free strained silicon-on silicon by in-place bonding. Appl. Phys. Lett. **86,** 251902 (2005).

33. Owen, D. L., Lackner, D., Pitts, O., Watkins, S. & Mooney, P. Bonding of Elastically strain-relaxed GaAs/InGaAs/GaAs heterostructures to GaAs (001). ECS Transactions, **16,** 271-278 (2008).

34. Owen, D. L., Lackner, D., Pitts, O. J., Watkins, S. P. & Mooney, P. M. In-place bonding of GaAs/InGaAs/GaAs heterostructures to GaAs (001). Semicond. Sci. Technol. **24,** 035011 (2009).

35. Plumhof, J. D. *et al.* Experimental investigation and modeling of the fine structure spliltting of neutral excitons in strain-free GaAs/Al$_x$Ga$_{1-x}$As quantum dots. Phys. Rev. B **81,** 121309 (2010).

36. Stevenson, R. M. *et al.* Strong directional dependence of single-quantum-dot fine structure. Appl. Phys. Lett. **87,** 133120 (2005).

37. Bir, G. L. & Pikus, G. E. *Symmetry and Strain-induced Effects in Semiconductors*. (John Wiley & Sons, New York, 1974).

38. Bester, G. Electronic excitations in nanostructures: an empirical pseudopotential based approach. J. Phys.: Cond. Matter **21**, 023202 (2009).

39. Bester, G., Nair, S., & Zunger, A. Pseudopotential calculation of the excitonic fine structure of million-atom self-assembled In$_{1-x}$Ga$_x$As/GaAs quantum dots. Phys. Rev. B 67, 161306(R) (2003).

40. Rastelli, A. *et al*. Controlling quantum dot emission by integration of semiconductor nanomembranes onto piezoelectric actuators. Phys. Stat. Sol. (B) **249,** 687-696 (2012).

41. Trotta, R. *et al.* Nanomembrane quantum-light-emitting diodes integrated onto piezoelectric actuators. Adv. Mater. **24,** 2668-2672 (2012).

42. R. Trotta, *et al*. Universal recovery of the energy-level degeneracy of the bright excitons in InGaAs quantum dots without a structure symmetry. Phys. Rev. Lett. (in press).

43. Bayer, M. *et al.* Fine structure of neutral and charged excitons in self-assembled In(Ga)As/(Al)GaAs quantum dots. Phys. Rev. B **65,** 195315 (2002).

44. Tonin, C. *et al.* Polarization properties of excitonic qubits in single self-assembled quantum dots. Phys. Rev. B **85,** 155303 (2012).